# Anisotropic Anomalous Diffusion assessed in the human brain by scalar invariant indices

S.De Santis*[a,b] , A.Gabrielli[a,c] , M.Bozzali[b] , B.Maraviglia[a,b], E.Macaluso[b],
S.Capuani[a,d]

[a] Physics Department, Sapienza University, P.le A.Moro,5 00185 Rome, Italy

[b] Neuroimaging Laboratory, S. Lucia Foundation, Via Ardeatina, 306 00179 Rome, Italy

[c] ISC-CNR, via dei Taurini, 19 00185 Rome, Italy

[d] CNR IPCF UOS Roma Sapienza, Physics Department, Sapienza University, P.le A.Moro,5 00185 Rome, Italy

∗ Corresponding author: Silvia De Santis

Physics Department, Sapienza University, P.le A.Moro,5 00185 Rome, Italy

Telephone: +390649913928

Fax: +390649913928

Email address: silvia.desantis@roma1.infn.it (S.De Santis)

**Abstract**

A new method to investigate anomalous diffusion in human brain is proposed. The method was inspired by both the stretched-exponential model proposed by Hall and Barrick (HB) and DTI. HB quantities were able to discriminate different cerebral tissues on the basis of their complexity, expressed by the stretching exponent $\gamma$ and by its anisotropy across different directions. Nevertheless, these indices were not defined as scalar invariants. In the present work, the signal was expressed as a simple stretched-exponential only along the principal axes of diffusion, while in a generic direction it was modeled as a combination of three different stretched-exponentials. In this way, indices to quantify both the tissue anomalous diffusion and its anisotropy, independently of the reference frame of the experiment, were derived. The method was tested and compared with DTI and HB approaches on 10 healthy subjects at 3T. The experimental results show that the new parameters are highly correlated to intrinsic local geometry when compared to HB indices. Moreover, they offer a different contrast when compared to DTI outputs. Specifically, the new indices show a higher capability to discriminate among different areas of the corpus callosum, which are associated to different axonal densities.

## Introduction

Diffusion Weighted Magnetic Resonance Imaging (DWI) enables the diffusional motion of water molecules to be measured, providing a unique form of contrast among tissues. In fact, due to the interactions between water molecules and cellular structures, DWI provides information about the size, shape, orientation and geometry of biological tissues. In living tissues like the human brain, the diffusion coefficient is generally dependent upon the direction along which it is measured; that is, it is anisotropic. Such anisotropy reflects, to some extent, the underlying fiber structure (1). This observation prompted the development of Diffusion Tensor Imaging (DTI) (2,3). DTI is based on the diffusion tensor reconstruction, which is obtained by combining diffusion measurements in at least six non-collinear spatial directions. In order to characterize the orientation-dependent water mobility in each voxel and to correlate it with the tissue architecture, parametric maps are usually displayed. Indices such as the mean diffusivity (MD) and the degree of anisotropy (FA) of the media, obtained from the diffusion tensor, provide information about underlying microstructural characteristics of biological tissues.

The central nervous system includes at least three different compartments: the gray (GM), the white matter (WM), and the cerebrospinal fluid (CSF). In the CSF, water molecules are in a barrier-free environment, which is characterized by unrestricted diffusion and no preferential directions. Conversely, WM and GM are characterized by a structural complexity which hinders, at different degrees, the water mobility, thus reducing the MD values. Moreover, WM is the highest structured cerebral tissue since it contains bundles of nerve fibers, which act like obstacles and traps for the diffusing water molecules. The presence of axons also creates privileged diffusive patterns and an anisotropic diffusion, which results in an increase of FA. For all these reasons, DTI has been largely used to investigate subtle abnormalities occurring in a variety of neurological diseases and psychiatric disorders (4-8).

In DTI acquisitions, the signal is typically recorded by diffusion-sensitized sequences as a function of chosen b-values, and it can be described as a monoexponential decay using the Stejskal-Tanner equation (9). Nonetheless, in the last few years, several experiments have demonstrated that models based on a single exponential curve have poor data predictability. This evidence comes from a number of studies on both animal models (10) and humans (11-13). Several approaches have been suggested to give a deeper insight into the diffusive phenomenon, in order to identify a better agreement between the data and the proposed fitting curves (14-18).

An innovative strategy was introduced by Bennett et al. (19). The signal decay as a function of the b-value, was modeled as a stretched exponential where the stretching exponent $\gamma$ was linked to the heterogeneity of the media in which spins diffusion occurs. The method was applied to the healthy human brain (19, 20) showing the ability to discriminate between different tissues on the basis of their structural complexity. Moreover, this approach has also been used for the investigation of brain tumors in animal models (21) as well as in humans (22, 23), showing promising results in terms of image contrast.

In the framework of the stretched exponential model, it might be of interest to investigate anomalous diffusion in the presence of anisotropic environments, which are known to characterize the central nervous system. So far, no formal theoretical models have been proposed to take into account the anisotropy of the stretching exponent. Recently, a first important experimental approach was introduced (20). These Authors developed a method to quantify not only the magnitude of the stretching exponent, but also its anisotropy. Their approach was based on the measurement of the so called anomalous exponent γ across several gradient directions, thus obtaining the mean anomalous exponent (AE) and its spread along different directions, i.e., the anomalous anisotropy (AA). According to Hall and Barrick, AA should be regarded as an equivalent of the FA. However, while FA quantifies the mean squared gap between three eigenvalues of the diffusion tensor and their mean value, AA estimates the difference between the stretching exponents as measured in *n* different directions and their correspondent mean values. The main difference between the outputs of the DTI analysis and the anomalous exponent indices is that, while the former are defined as scalar invariants of a tensor (i.e. the diffusion tensor) and are by definition independent of the reference frame in which they are measured, the latter instead depend on the directions along which the average is quantified.

We propose here an alternative method to account for the dependence of the stretching exponent from the spatial direction. Using an approach similar to that for deriving the anisotropic diffusion from tensor calculation, we assume the stretched exponential model to be valid along three principal directions (i.e. the main axes of diffusivity) only, rather than along *n* arbitrary directions. Indeed, in the three-dimensional space, the motion can always be expressed by a combination of three components, which depend on the geometry of the local system. If the measurement is performed along one of these main directions, the decay can be expressed as a simple stretched exponential. As a consequence, when the measurement is performed along a generic direction, it is reasonable to model the signal as the superimposition of the decay along each of the main directions, thus involving all the three main exponents.

By repeating the measurement along several directions and performing simultaneous fits, it is therefore possible to obtain three main exponents and their corresponding anisotropy factors. As a consequence, the results are independent of the laboratory reference frame. Using this strategy, we are able to quantify the mean values of the anomalous exponents and their anisotropy, which we will define as Mγ and γA respectively throughout the manuscript. These indices are thus similar to those defined in (20), but the dependence on the laboratory frame has been removed. For these reasons, Mγ and γA are scalar invariant indices which can be used to assess anomalous diffusion in tissues.

Specific aims of the current work were: 1) to define a new procedure to obtain *in vivo* Mγ and γA; 2) to compare these measures with MD and FA and with AE and AA as defined by Hall and Barrick (20). We will show that γA is more closely correlated with FA than AA. Moreover, we will demonstrate that Mγ reflects aspects of the diffusive dynamics which are different from those caught by MD. For these reasons, our approach to anomalous diffusion may offer new chances to improve the brain tissue characterization.

## Methods

*Theory*

The DTI theory (2, 3) assumes that the signal which is recorded by a diffusion-sensitized sequence may be expressed as a mono-exponential decay, according to the following equation (9):

$$S(b) = S(0)\exp(-Db) \qquad [1]$$

where $D$ is the apparent diffusion constant and $b$ is the so called b-value, i.e. the scalar value of gradient weighting strength. In three-dimentional space, Eq.[1] holds along the three main axes of diffusion which coincide with the principal self-diffusivity directions, with in general a specific value of D for each of these three directions. Along a generic direction, correlations between molecular displacements in perpendicular planes affect the signal decay, which has to take into account the coupling of the non-diagonal terms in the diffusion tensor (24, 25). Nevertheless, due to the special properties of the Gaussian distribution that characterizes the spins displacements, the signal does not lose its simple exponential form. In this framework, the scalar invariants MD and FA are derived from a specific rearrangement of diffusion tensor eigenvalues. The assumption underlying this approach is the linear relationship between mean-squared displacement of diffusing spins and diffusion time which is known to hold in homogeneous environments, according to the following equation:

$$\langle r^2(t)\rangle \propto Dt \qquad [2]$$

However, in non locally homogeneous environments such as porous or fractal media, which are characterized by the presence of obstacles, inhomogeneities and traps on many length scales (for a review, see (26)), the relationship between mean-squared displacements and diffusion time can no longer be expressed in a linear form as in Eq. [2]. In these cases, one or more stretching exponents appear, which quantify the deviation from the ideal conditions. In the one dimensional case (or equivalently in the isotropic d-dimensional one), the following equation is verified:

$$\langle r^2(t)\rangle \propto Dt^{\gamma} \qquad [3]$$

Where $\gamma$ is an adimensional index for which $0<\gamma\leq 1$. The effects of a non-linear dependence between mean-squared displacement and time modify the measured signal expression in a non-trivial way. Several Authors (20, 21, 23) have recently proposed the following stretched exponential form:

$$S(b) = S(0)\exp(-Db^{\gamma}) \qquad [4]$$

Hall and Barrick developed an approach which considers the anisotropy of $\gamma$. In their protocol, based on the acquisition of diffusion weighted images at different b-values, these Authors measured the direction-dependent exponent γ pixel-wise along several gradient directions. In this way, they obtained parametric maps of the mean value of the anomalous exponent (AE), i.e.

$$AE = \langle \gamma \rangle = \frac{1}{N}\sum_{i=1}^{N} \gamma_i \qquad [5]$$

and of the anomalous anisotropy (AA), i.e.

$$AA = \sqrt{\frac{N}{N-1}\frac{\sum_{i=1}^{N}\left(\gamma_i - \langle \gamma \rangle\right)^2}{\sum_{i=1}^{N}\gamma_i^2}} \qquad [6]$$

where $N$ is the number of the directions, $\gamma_i$ is the anomalous exponent measured in the i-th direction and $\langle\gamma\rangle$ is the mean exponent. However, as already mentioned in the introduction, AE and AA are not defined as scalar invariants. As a consequence, their value depends on the set of gradient directions chosen for the experiment.

We propose here a different approach to account for the γ anisotropy, which is similar to that used for FA estimation in DTI analysis. Our method aims at accounting for the tensorial nature of the anomalous diffusion, with a description which does not depend on the laboratory reference frame but which is intrinsic of the system. For this purpose, we draw an analogy with the ordinary diffusion dynamics, where the diffusion tensor is diagonal only along three main directions. In this case, Eq. [4] is valid along each of the three principal diffusive directions. When considering a generic direction, we hypothesize the total signal to be due to a combination of the behaviors along each of the three main axes, according to the following formula:

$$S(b) \propto \prod_{i=1}^{3} \exp\left(-A_i b^{\gamma_i}\right) \qquad [7]$$

where $A_i$ is a generalization of the diffusion constant and the b-value is calculated along the chosen measurement direction in the reference frame of the principal axes (i.e. it contains the director cosines of the measurement direction with respect to the principal reference frame). See Fig. [1] for a visual example of the method.

It is not possible to know *a priori* the directions associated to the principal diffusion axes, which are thereby linked to the local geometrical structure and therefore supposed to be voxel-dependent. In principle, the complete solution for this problem requires the estimation of 12 parameters: 3 for the $A_i$, 3 for the $\gamma_i$ and 6 to define the principal reference frame. We simplified this model by separating the analysis into two steps: we first calculate the principal reference frame by using the DTI analysis and then we evaluate the remaining six parameters by measuring the signal in at least six non collinear directions. This implies as an approximation that the principal reference frame is the same for both DTI and anomalous diffusion framework.

We will refer to the indices derived with our method as Mγ for the mean value of $\gamma_i$ and γA for its anisotropy calculated by following formula [6].

*Data Acquisition*

Ten healthy volunteers (F/M=4/6, mean age and standard deviation 24 ± 3 years) participated in this study after giving informed consent, according to the national laws and to the local ethics committee guidelines. All imaging was obtained using a head-only 3.0T scanner (Siemens Magnetom Allegra, Siemens Medical Solutions, Erlangen, Germany), equipped with a circularly polarized transmit-receive coil. The maximum gradient strength is 40 mT $m^{-1}$, with a maximum slew rate of 400 mT $m^{-1}$ $ms^{-1}$. T1-weighted sagittal images were acquired for anatomical reference and brain segmentation, using a MPRAGE (Magnetization-Prepared Rapid Acquisition with Gradient Echo sequence) [TR/TE/TI=2000/4.38/910 ms, flip angle=8°, matrix=448×512, in-plane resolution=0.5×0.5 $mm^2$, slice thickness=1 mm, 176 adjacent slices, field of view (FOV) of 224×256 $mm^2$]. Diffusion weighted (DW) SE EPI were acquired to cover the whole brain using the following parameters: TR/TE= 6400/107 ms, Δ/δ=107/35ms, bandwidth=1860 Hz/px, slice thickness=3mm, in plane resolution=1.8×1.8 $mm^2$. The encoding gradients were applied along 6 non collinear directions at 16 different b-values: 0, 100, 200, 300, 400, 500, 700, 800, 1000, 1200, 1500, 2000, 2400, 3000, 4000, 5000 s/$mm^2$.

*Data Analysis*

All DW images were corrected for eddy currents distortions using FSL version 4 (http://www.fmrib.ox.ac.uk/fsl) software. As a first step, FA and MD were derived after tensor calculation by means of FSL DTIFIT routine using *b*=1000 s/$mm^2$. Besides, the three eigenvectors which define pixel-wise the principal reference frame, were obtained. Anatomical scans were then co-registered to FA maps and segmented into white matter (WM), grey matter (GM) and Cerebrospinal Fluid (CSF) using SPM version 5 (http://www.fil.ion.ucl.ac.uk/spm). A binary mask was obtained by combining WM, GM and CSF segments and retaining only voxels with intensity greater than 0.8 on the resulting image. A custom script (Matlab, The Mathworks, Natick, MA, USA) was implemented to perform the image reconstruction. The $\gamma_i$ values were obtained by means of a non-linear least-squares (utilizing Levemberg-Marquardt minimization) multi-dimensional estimation procedure of formula [7], in the subspace of *b* and $A_i$.

Parametric maps based on Mγ and on γA were then obtained. To compare the results with the anomalous exponent analysis proposed by Hall and Barrick, AE and AA were also derived by means of a custom script, written following the prescriptions found in the published paper (20).

MD, FA, Mγ, γA, AE and AA were measured in 13 regions of interest (ROIs) selected on the $b_0$ images as previously described (27). Data have not been normalized into a stereotaxic space but kept in their own native one, so the ROIs were placed manually for every subject. Rectangular ROIs of variable volume (range 68.4-420.3 $mm^3$), depending on the anatomical region studied, were placed bilaterally in the following areas: the occipital(a) and temporal lobe(b), the anterior pericallosal areas(c), the genu(d) and the splenium(e) of the corpus callosum, the posterior pericallosal areas(f),the frontal(g) and parietal lobe(h), the thalamus(i), the putamen(l), the head of the caudate nucleus(m), the posterior limb and the genu of the internal capsule(n). An additional control ROI was defined within each lateral ventricle (o). Figure 2 illustrates the location of all parenchymal ROIs. The ROIs were eventually transferred onto all considered quantitative maps for each subject, and average measures were calculated for every ROI. Two kind of statistical analysis were performed to evaluate our method. First of all, Pearson correlation coefficient (r) was obtained between FA, γA and AA and between MD, Mγ and AE. Besides, mean and SD were obtained in each ROI and ANOVA was used to test the efficacy of our parameters in discriminating between considered regions, compared to DTI. As a first step, Lilliefors test was performed to confirm the normality of the distribution across subjects. Then we performed a two-way ANOVA, in which the two factors were the regions (12 levels) and the methods used (two levels, DTI or our approach). The ANOVA was performed separately for the methods FA-γA and for MD-Mγ. Following significant interaction in the ANOVA for both FA-γA and MD-Mγ, post-hoc t-tests were performed for each couple of regions.

**Results**

Figure 3 shows an example of MD, FA, Mγ, γA, AE and AA maps obtained from one studied subject. A different contrast was evident between MD/FA and anomalous diffusion maps (Mγ, γA, AE and AA). Interestingly, within the framework of the anomalous diffusion model, the two analyses (i.e. Mγ, γA vs AE, AA) lead to different contrasts. From a visual inspection of the images, in both MD and FA and our anomalous diffusion maps Mγ and γA there are present anatomical landmarks which are not visible in AE and AA maps.

As a first step of the data analysis, our specific aim was to investigate the correlations between DTI indices and anomalous diffusion parameters, derived with both methods. In Fig.4A and 4B there are reported the correlation plots across ROIs between FA and γA and between FA and AA. The plot 4A clearly shows that there is a high positive correlation between FA and γA ($r$=0.91, $p$<0.0001 without CSF, $r$=0.92, $p$<0.0001 with CSF). Conversely, the linear correlation between FA and AA is less remarkable ($r$=0.47, $p$=0.063 without CSF, $r$=0.54, $p$=0.028 with CSF), and the error bars associated to data points overlap in the AA axis, thus suggesting that poor additional information can be gathered by estimation of AA.

The correlation plots across ROIs between MD and Mγ and between MD and AE are shown in Fig.4C and 4D respectively. The resulting trend between MD and Mγ is positive ($r$=0.45, $p$=0.069 without CSF, $r$=0.92, $p$<0.0001 with CSF) but moderate, especially when the CSF contribution is not considered. It is noticeable that some regions are clearly discriminated by Mγ, while their MD values overlap when the error bars are considered. For example, in the corpus callosum the splenium has a Mγ value significantly lower than the genu, as shown in Fig.4C. Conversely, MD and AE are poorly correlated to each other ($r$=0.34, $p$=0.14 without CSF, $r$=0.39, $p$=0.096 with CSF).

Since the correlation plots were calculated between the mean values of MD, FA, Mγ, γA, AE and AA, averaged across all subjects, to investigate if those correlations are significant (i.e. found also in each single subject) or if they are merely an effect of the average, we calculated the correlation coefficient for each of the ten subjects. In Fig.5 there is reported the correlation coefficient for MD and Mγ, MD and AE, FA and γA, FA and AA. The correlation coefficient which relates the DTI indices with Mγ and γA is always positive and larger than 0.6, showing that the positive correlations reported in Fig.4A and 4C reflects the trend found at single subject level. Conversely, the correlation coefficient calculated between MD and AE and between FA and AA fluctuates between positive and negative values. This means that the moderate correlation which is reported in Fig.4B and 4D is not found in every single subject but it is just a consequence of the values averaging.

In the second part of the study we evaluated the capability of our new indices Mγ and γA to discriminate between different cerebral tissue, compared to DTI indices. In Fig.6 there are reported the mean values and SD averaged across all subjects into the considered ROIs, of FA, γA, AA, MD, Mγ and AE. In the histogram 6A there are reported FA values (left), γA and AA (right), all in adimensional units. In the histogram 6B there are reported MD values (in $m^2/s$, left), Mγ and AE (in adimensional units, right).

In order to investigate the ability in discriminating among different cerebral regions of FA compared to γA and of Mγ compared to MD, two-way ANOVA tests were performed. The F-value associated to the levels FA-γA was F(11)=70 while the F-value associated to the levels MD-Mγ was F(11)=34. Following significant interaction in the ANOVA, paired t-tests are reported in Tab.1A for FA, Tab.1B for γA, Tab.2A for MD and Tab.2B for Mγ. By comparing Tab.1A with Tab.1B, it is evident that most of the regions which are highly discriminated ($P$<0.001, in gray) by γA are highly discriminated by FA as well. Moreover, FA is able to highly discriminate some regions which are not discriminated or moderately discriminated by γA, i.e. FA values have more discriminating power compared to γA. An exception to this trend is represented by the highest anisotropic structures, i.e. the genu and the splenium of the corpus callosum, which instead are better discriminated by γA than by FA.

Conversely, several regions which are not statistically discriminated ($P>0.05$) by conventional MD, turned out to be discriminated on the basis of Mγ. As an example, the splenium is always associated to a $P_{M\gamma}<0.05$ when correlated to each one of the other ROIs (see Tab.2A), while $P_{MD}>0.05$, except for the putamen, the occipital and temporal lobe (see Tab.2B). Besides, couples of regions associated to highly significant p-values ($P<0.001$, in gray) on Mγ basis are different to those highly discriminated by MD. For example, the genu and the splenium are discriminated by MD with a p-value of 0.05 but are better discriminated by Mγ with $P<0.001$.

**Discussion**

In the current study, anomalous diffusion maps, Mγ and γA, obtained by considering the stretched exponential model across the three main diffusivity axes were compared with both MD and FA maps, and with the anomalous diffusion indices AE and AA, obtained using the model recently proposed by Hall and Barrick.

A first result is that the cerebral tissue is anisotropic also respect to γ. Bennett and coworkers (19) introduced the stretched exponential model as a fitting function to obtain maps of the γ exponent across one selected direction only, in analogy with DWI. The sensitivity to the chosen direction was further tested by the same group (28), showing that the stretching exponent is insensitive to the orientation of the applied magnetic field gradient. These Authors measured the stretching exponent across three orthogonal directions reporting a small anisotropy which was roughly constant in both GM and WM tissues. Conversely, other Authors (20) reported a difference between GM, WM and CSF with respect to their anisotropy in γ measured across 12 different directions. In this regard, our work confirmed that there exists an anisotropy in the stretched exponent, on whose basis we were able to discriminate between different cerebral structures.

The high correlation we found between γA and FA confirms that in the framework of the anomalous diffusion, the definition of three main diffusivity directions may be a good strategy to obtain a quantification of anisotropy which is independent of the reference frame. Their high correlation indicates that both quantities, i.e. FA and γA refer to intrinsic geometrical properties of brain tissues, which are independent of the reference frame in which the gradient directions are expressed. Moreover, post-hoc t-tests highlighted that the highest anisotropic structures, i.e. the genu and the splenium of the corpus callosum, are better discriminated by γA than by FA.

Conversely, the anisotropy index obtained by considering each direction as characterized by a single stretched exponential decay, i.e. AA, revealed a poor correlation with FA, which fluctuates between positive and negative values at single-subject level. The low correlation found between AA and FA, i.e.

between two measures which are supposed to be dependent on the geometry, confirm that Hall and Barrick's method suffers from the dependence from the reference frame in which the measurement is performed. Nevertheless, if the number of chosen directions is enough to sample the space uniformly, then these effects are likely to be smoothed. To reduce the experimental time, we acquired data using gradients applied along 6 directions only, which is the minimal number required to perform DTI calculation. Conversely, in the work published by Hall and Barrick, 12 different directions were chosen. Even though these Authors did not show any correlation plot, the contrast-to-noise ratio of the obtained maps seems to be higher as compared to our results, thus confirming the relevance of the number of directions. Nevertheless, the current study underlines the limitation of Hall and Barrick's approach and highlights the importance of employing an intrinsic method to quantify anomalous diffusion indices.

On the other hand, one of the main limitations of our method is its approximation in considering the principal directions of diffusion as the same obtained using the DTI model. This may have enhanced the resulting correlations that we found between FA and γA. However, this approximation is reasonable. In fact, since the magnitude of $\gamma$ is always slightly lower than 1 (ranging from 0.7 to 1, where for $\gamma=1$ the Stejskal-Tanner mono-exponential decay holds), we expect only a small difference in the spatial orientation of the two reference frames. Moreover, we expect this approximation to hold if the main diffusive axes are independent of the b-value. DTI reconstruction is in fact performed at a relatively low b-value of 1000 s/mm$^2$ while the anomalous diffusion method uses higher b-values (up to $b$=5000 s/mm$^2$). This issue is crucial since exploring higher b-values means probing slower dynamics which can be in principle linked to different spatial arrangements. Nevertheless, to our knowledge, diffusion models which take into account the different diffusion pools proposed in literature consider the same principal axes for both, slower and faster diffusion subgroups (29-31). As a consequence, we hypothesize here that when the b-value range is extended from 1000s/mm to 5000s/mm the eigenvectors of the diffusion tensor remain unchanged or change only slightly. In order to overcome this approximation, a future work is needed in which more gradient directions are selected. In this case, we would be able to evaluate, in the multi-dimensional fit, not only the stretching exponents and their relative amplitude factors, but also the director cosines associated with the main diffusion axes, thus avoiding to assume any *a priori* information about the principal reference frame orientation.

The results concerning Mγ need further studies and a future validation. The lack of a high correlation between MD and Mγ is encouraging because it suggests that the two measures provide a different structural information. That is to say, there exist regions in which the diffusion is restricted but not anomalous. The two properties indeed correspond to different physical phenomena. The restricted diffusion is due to barriers which constrain the water molecules motion inside a portion of the space which is smaller than that travelled if the environment is barrier-free, as reflected by a reduction of the MD (32).

Conversely, the anomalous diffusion is associated to the complexity of the path travelled by the spins, which depends on the shape and size distribution of the barriers. For example, it has been demonstrated that neurons have a fractal-like appearance (23, 33-35). As a consequence, two situations characterized by the same average mean free path but different barriers distributions might be better discriminated by $M\gamma$, as confirmed by comparing the post-hoc t-tests reported in Tab.2. In this regard, we found an interesting difference in $M\gamma$ values associated to two distinct areas of the corpus callosum, i.e. the genu and the splenium, which are instead overlapped in the MD axis in our study (Fig.4C) and for which also in literature similar diffusivity values are reported (36, 37). Since the stretching exponent is postulated to be sensitive to the presence of traps and obstacles on many different length scales, we can speculate that a broader distribution of axonal diameters would result in a lower $M\gamma$. A number of publications underlined the presence of uneven distributions of fiber types along the corpus callosum (38). A recent work introduced an powerful method to evaluate the axon diameter distribution by means of diffusion MRI (39). As this method was applied *in vivo* to the corpus callosum of the rat brain (40), it showed different axonal density distributions which are moreover characterized by different widths. In particular, the genu was associated to a narrower distribution compared to the splenium, i.e. in the splenium different axonal diameters coexist. We speculate that this diameters heterogeneity can explain the difference in the $M\gamma$ that we observed between the splenium and the genu of the corpus callosum, as reported in Fig.4C.

**Conclusions**

In the framework of the anomalous diffusion, we propose here an innovative method of considering the spatial dependence of the stretching exponent. On the basis of Hall and Barrick's results, indices analogous to the MD and FA were derived, i.e. $M\gamma$ and $\gamma A$. To characterize these parameters, the correlation with DTI matrixes was explored and their specificity in discriminating between different cerebral structures was tested on ten healthy subjects. The high correlation between $\gamma A$ and FA demonstrates that our approach does not suffer from the dependence on the reference frame. Besides, $M\gamma$ proves to be able to reveal a different information when compared to MD, due to its capability of discriminating between specific cerebral regions which are not distinguishable on MD basis. For these reasons, our method is eligible for the characterization of the brain tissue. Besides, we believe that our analysis may provide a different contrast also when applied to the characterization of microstructural alterations, compared to DTI. For this reason, the next step will be to select specific neurological diseases for an *in vivo* application.

**Figure Captions**

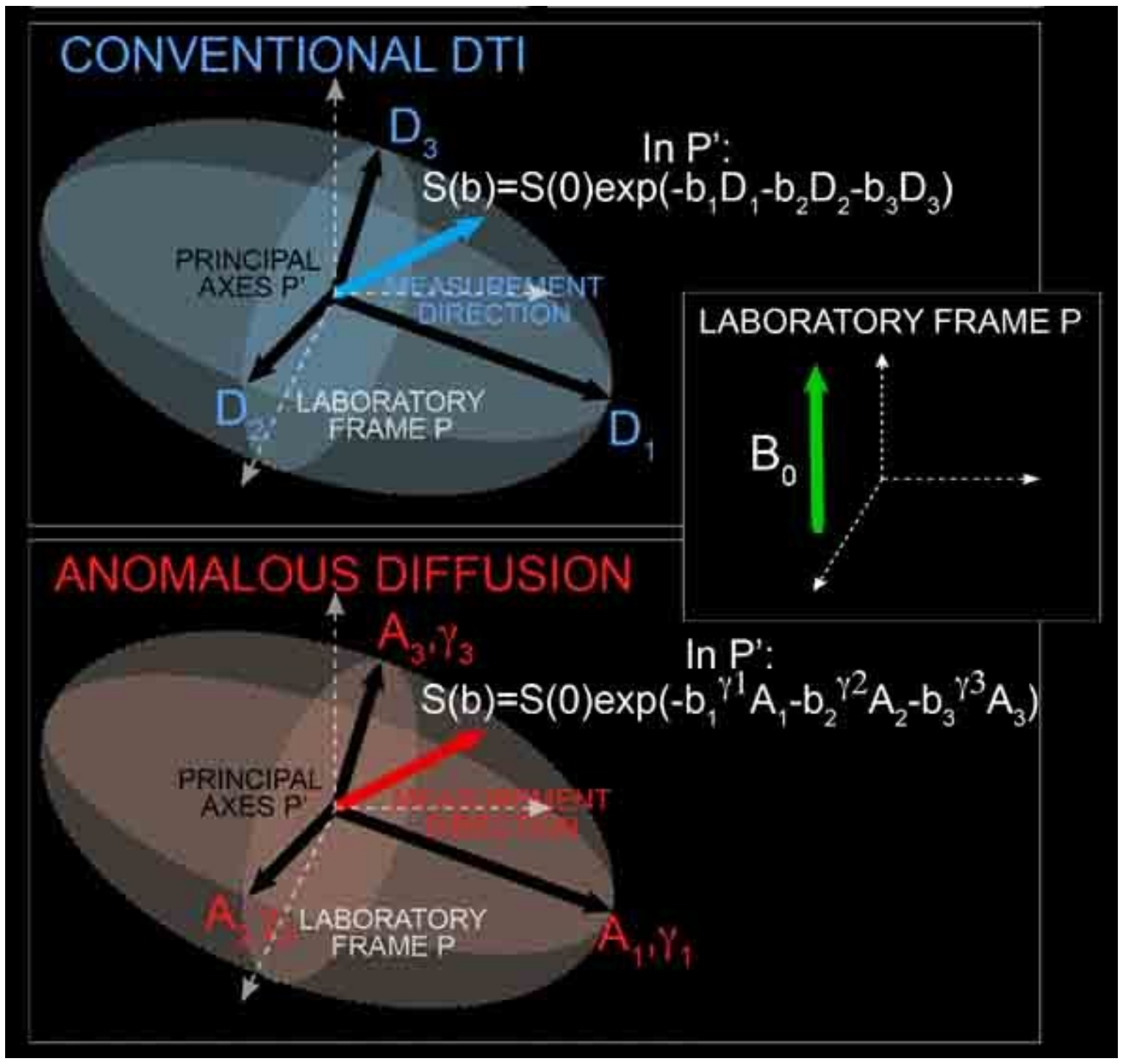

Figure 1: Visual example of the analogies between DTI and our approach to describe the anomalous diffusion. For more details on the method, see the theory paragraph.

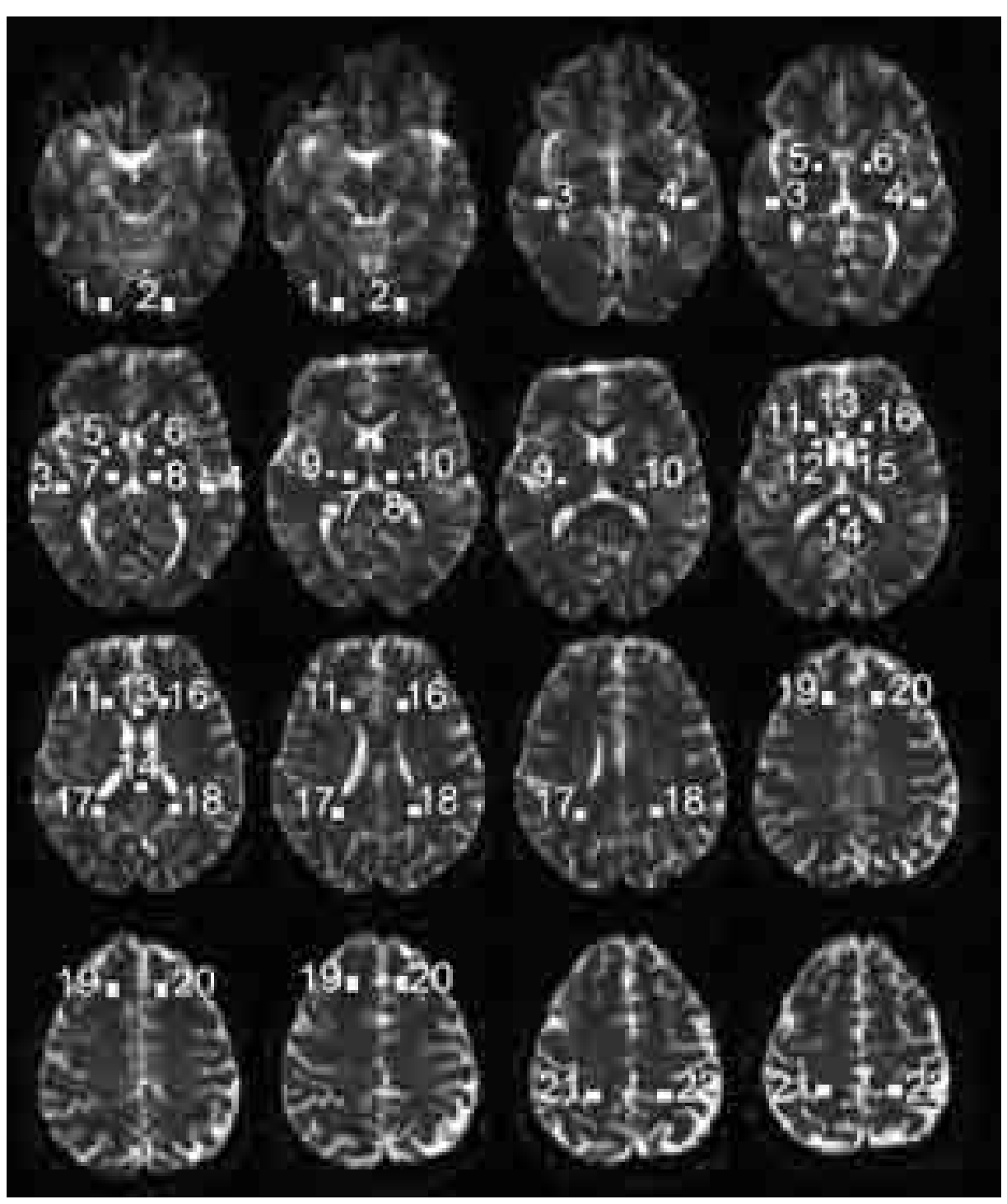

Figure 2: Location of the selected regions of interest (ROIs) in a healthy volunteer: b1-b2 temporal lobe, a1-a2 occipital lobe , l1-l2 putamen, i1-i2 thalamus, n1-n2 posterior limb of the internal capsule, d-e genu

and splenium of the corpus callosum, c1-c2 anterior pericallosal areas, f1-f2 posterior pericallosal areas, m1-m2 head of the caudate nucleus, g1-g2 frontal lobe, h1-h2 parietal lobe. All the ROIs are superimposed on B0 images.

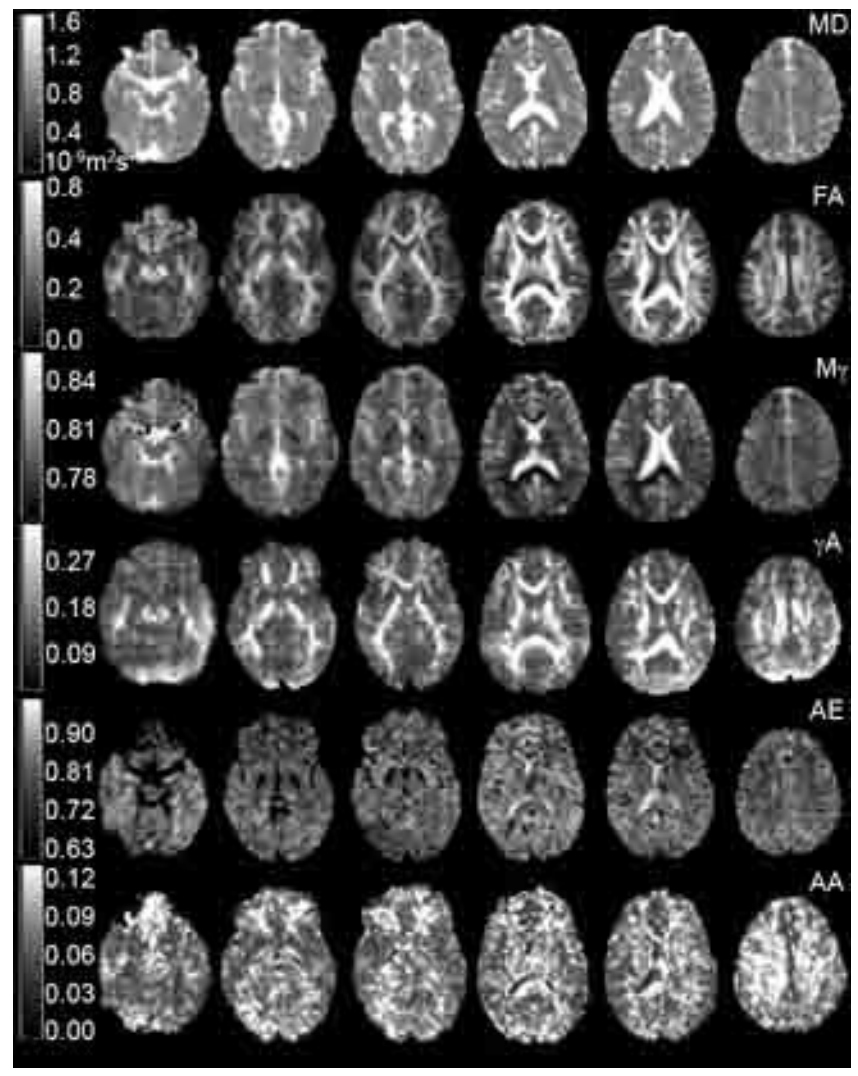

Figure 3: From top to bottom: mean diffusivity (MD), fractional anisotropy (FA), mean γ (Mγ), γ anisotropy (γA), mean anomalous exponent (AE) and anomalous anisotropy (AA) maps of the slices 3-12-16-23-27-36 from a healthy subject.

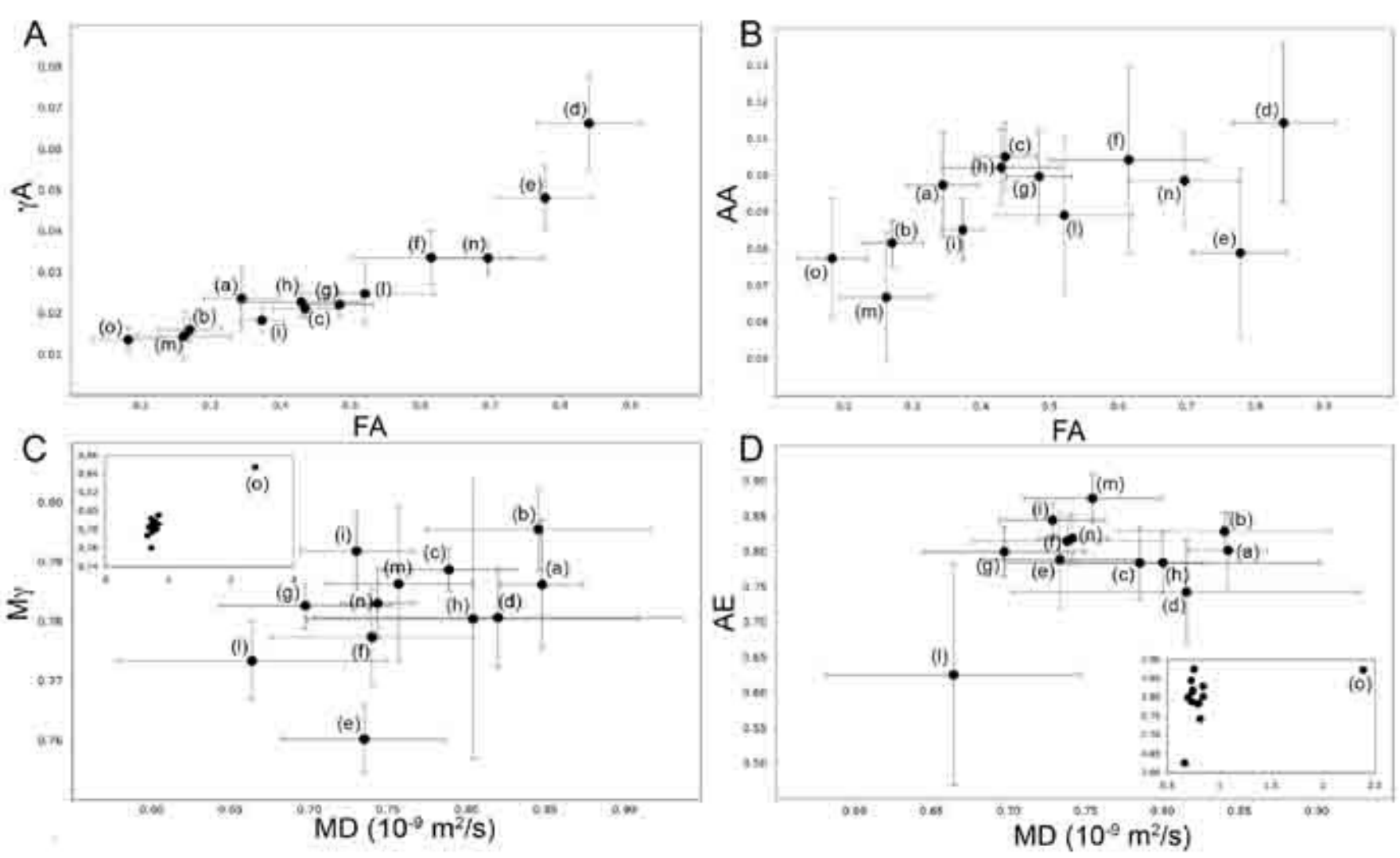

Figure 4: Correlation plots in which every point represents the value calculated for each of the 12 ROIs. Left: correlation between FA and γA (A) and between FA and AA (B). Right: correlation between MD and Mγ (C) and between MD and AE (D). Mean values were derived from the ROIs illustrated in Fig. 2: occipital lobe(a), temporal lobe(b), anterior pericallosal areas(c), genu(d) and splenium(e) of the corpus callosum, posterior pericallosal areas(f), frontal lobe(g), parietal lobe(h), thalamus(i), putamen(l), head of the caudate nucleus(m) and posterior limb of the internal capsule(n). In the inserts, the data point corresponding to the ROI drawn in the CSF(o) is added to the plot.

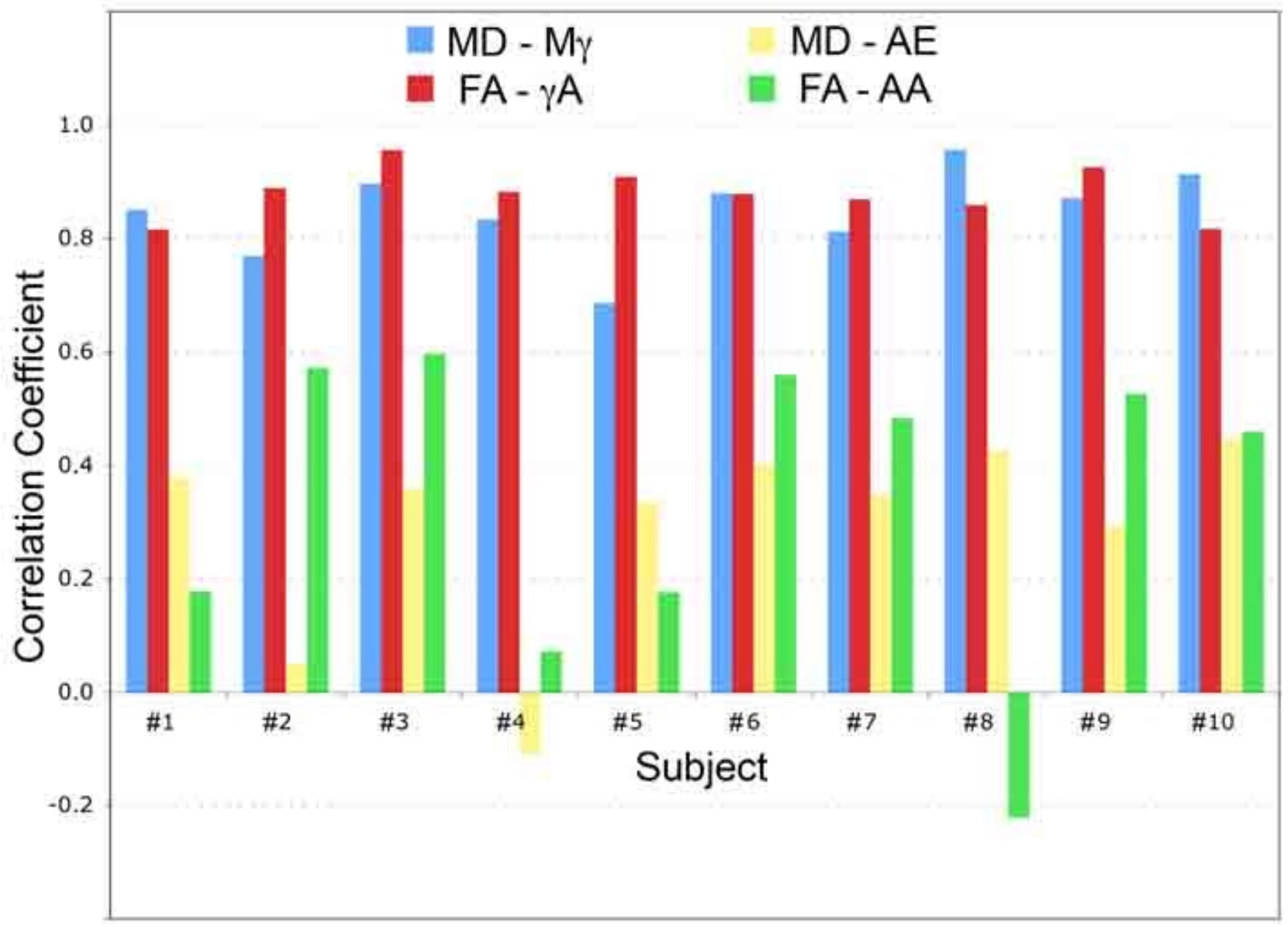

Figure 5: Correlation coefficient for mean diffusivity (MD) and mean γ (Mγ) (blue), fractional anosotropy (FA) and γ anisotropy (γA) (red), mean diffusivity (MD) and mean anomalous exponent (AE) (yellow), fractional anisotropy (FA) and anomalous anisotropy (AA) (green), plotted for each of the 10 subjects.

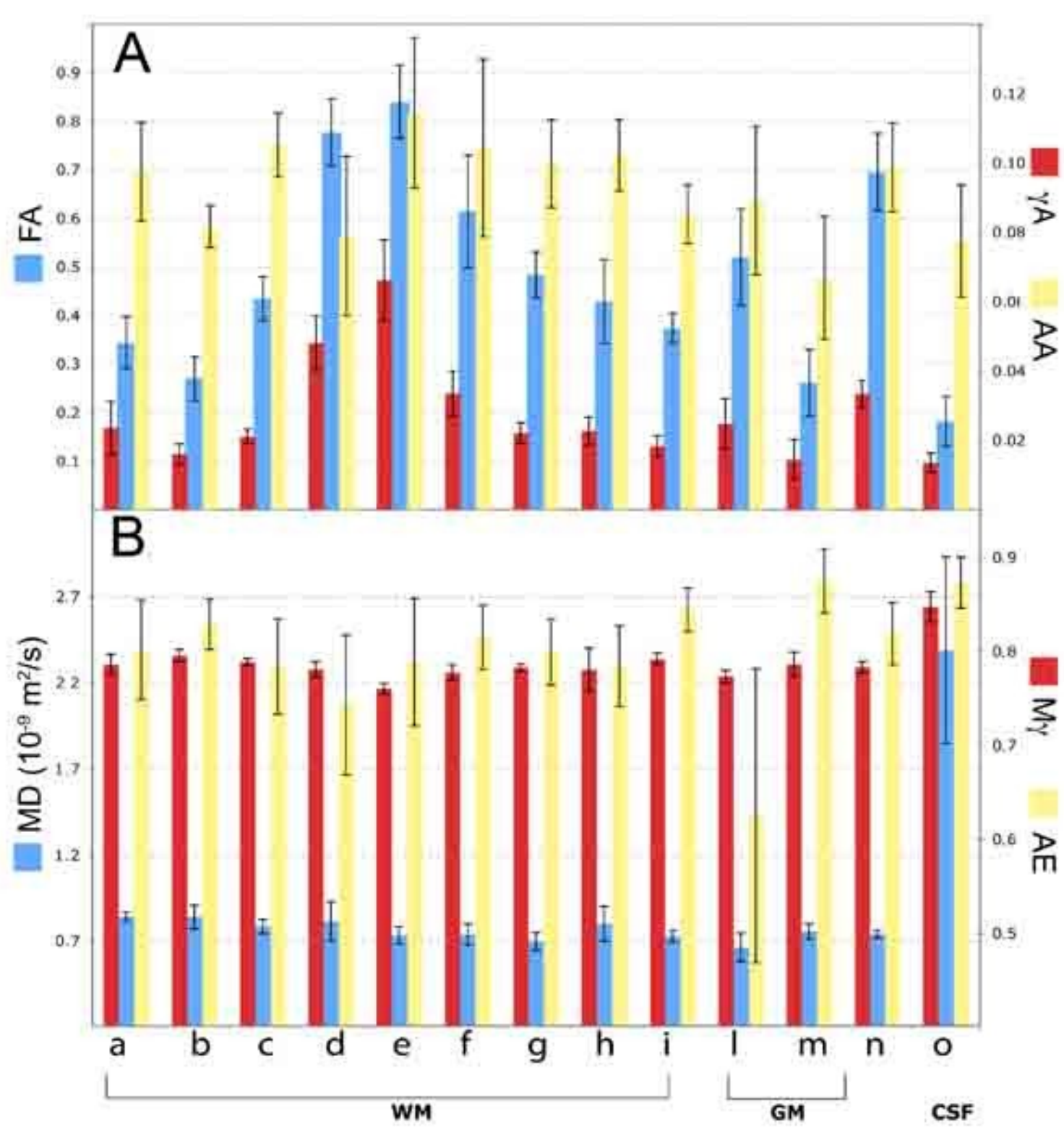

Figure 6: Mean values and associated standard deviations of fractional anisotropy (FA), γ anisotropy (γA),

anomalous anisotropy (AA) (A) and mean diffusivity (MD), mean $\gamma$ ($M\gamma$), mean anomalous exponent (AE) (B) derived from the ROIs illustrated in Fig. 2: occipital lobe(a), temporal lobe(b), anterior pericallosal areas(c), genu(d) and splenium(e) of the corpus callosum, posterior pericallosal areas(f), frontal lobe(g), parietal lobe(h), thalamus(i), putamen(l), head of the caudate nucleus(m), posterior limb of the internal capsule(n) and ventricle(o).

## Tables

| A | a | b | c | d | e | f | g | h | i | l | m | n |
|---|---|---|---|---|---|---|---|---|---|---|---|---|
| a | - | | | | | | | | | | | |
| b | 0.004 | - | | | | | | | | | | |
| c | P<0.001 | P<0.001 | - | | | | | | | | | |
| d | P<0.001 | P<0.001 | P<0.001 | - | | | | | | | | |
| e | P<0.001 | P<0.001 | P<0.001 | 0.067 | - | | | | | | | |
| f | P<0.001 | P<0.001 | P<0.001 | 0.001 | P<0.001 | - | | | | | | |
| g | P<0.001 | P<0.001 | 0.029 | P<0.001 | P<0.001 | 0.014 | - | | | | | |
| h | 0.016 | P<0.001 | 0.854 | P<0.001 | P<0.001 | P<0.001 | 0.104 | - | | | | |
| i | 0.191 | P<0.001 | 0.005 | P<0.001 | P<0.001 | P<0.001 | P<0.001 | 0.103 | - | | | |
| l | P<0.001 | P<0.001 | 0.023 | P<0.001 | P<0.001 | 0.066 | 0.307 | 0.041 | 0.005 | - | | |
| m | 0.008 | 0.774 | P<0.001 | P<0.001 | P<0.001 | P<0.001 | P<0.001 | P<0.001 | P<0.001 | P<0.001 | - | |
| n | P<0.001 | P<0.001 | P<0.001 | 0.025 | 0.001 | 0.085 | P<0.001 | P<0.001 | P<0.001 | P<0.001 | P<0.001 | - |
| | | | | | | | | | | | | |
| **B** | a | b | c | d | e | f | g | h | i | l | m | n |
| a | - | | | | | | | | | | | |
| b | 0.010 | - | | | | | | | | | | |
| c | 0.850 | P<0.001 | - | | | | | | | | | |
| d | P<0.001 | P<0.001 | P<0.001 | - | | | | | | | | |
| e | P<0.001 | P<0.001 | P<0.001 | P<0.001 | - | | | | | | | |
| f | 0.006 | P<0.001 | P<0.001 | P<0.001 | P<0.001 | - | | | | | | |
| g | 0.590 | P<0.001 | 0.395 | P<0.001 | P<0.001 | P<0.001 | - | | | | | |
| h | 0.760 | P<0.001 | 0.571 | P<0.001 | P<0.001 | P<0.001 | 0.717 | - | | | | |
| i | 0.089 | 0.120 | 0.007 | P<0.001 | P<0.001 | P<0.001 | 0.015 | 0.020 | - | | | |
| l | 0.741 | 0.002 | 0.571 | P<0.001 | P<0.001 | 0.011 | 0.910 | 0.459 | 0.037 | - | | |
| m | 0.008 | 0.140 | 0.003 | P<0.001 | P<0.001 | P<0.001 | 0.001 | 0.002 | 0.108 | 0.003 | - | |
| n | 0.002 | P<0.001 | P<0.001 | P<0.001 | P<0.001 | 0.963 | P<0.001 | P<0.001 | P<0.001 | 0.004 | P<0.001 | - |

Table 1: Post-hoc t-tests for FA (A) and γA (B). For each couple of regions, there are reported the obtained p-values. The cells corresponding to regions discriminated with high significancy ($P<0.001$) are colored in gray. The statistical test was performed between the following selected ROIs: occipital lobe(a), temporal lobe(b), anterior pericallosal areas(c), genu(d) and splenium(e) of the corpus callosum, posterior pericallosal areas(f), frontal lobe(g), parietal lobe(h), thalamus(i), putamen(l), head of the caudate nucleus(m) and posterior limb of the internal capsule(n)

| A | a | b | c | d | e | f | g | h | i | l | m | n |
|---|---|---|---|---|---|---|---|---|---|---|---|---|
| a | - | | | | | | | | | | | |
| b | 0.922 | - | | | | | | | | | | |
| c | 0.001 | 0.044 | - | | | | | | | | | |
| d | 0.470 | 0.562 | 0.438 | - | | | | | | | | |
| e | P<0.001 | P<0.001 | 0.021 | 0.050 | - | | | | | | | |
| f | P<0.001 | 0.003 | 0.061 | 0.074 | 0.851 | - | | | | | | |
| g | P<0.001 | P<0.001 | P<0.001 | 0.008 | 0.133 | 0.129 | - | | | | | |
| h | 0.219 | 0.318 | 0.672 | 0.756 | 0.079 | 0.117 | 0.419 | - | | | | |
| i | P<0.001 | P<0.001 | 0.007 | 0.054 | 0.825 | 0.704 | 0.166 | 0.077 | - | | | |
| l | P<0.001 | P<0.001 | P<0.001 | 0.003 | 0.037 | 0.037 | 0.305 | 0.004 | 0.057 | - | | |
| m | P<0.001 | 0.004 | 0.128 | 0.131 | 0.333 | 0.508 | 0.018 | 0.211 | 0.199 | 0.007 | - | |
| n | P<0.001 | P<0.001 | 0.009 | 0.058 | 0.643 | 0.843 | 0.025 | 0.093 | 0.355 | 0.011 | 0.427 | - |

| B | a | b | c | d | e | f | g | h | i | l | m | n |
|---|---|---|---|---|---|---|---|---|---|---|---|---|
| a | - | | | | | | | | | | | |
| b | 0.035 | - | | | | | | | | | | |
| c | 0.495 | 0.013 | - | | | | | | | | | |
| d | 0.218 | P<0.001 | 0.012 | - | | | | | | | | |
| e | P<0.001 | P<0.001 | P<0.001 | P<0.001 | - | | | | | | | |
| f | 0.057 | P<0.001 | P<0.001 | 0.398 | P<0.001 | - | | | | | | |
| g | 0.344 | P<0.001 | 0.002 | 0.501 | P<0.001 | 0.083 | - | | | | | |
| h | 0.490 | 0.067 | 0.283 | 0.981 | 0.016 | 0.705 | 0.773 | - | | | | |
| i | 0.217 | 0.270 | 0.217 | 0.007 | P<0.001 | P<0.001 | 0.002 | 0.202 | - | | | |
| l | 0.005 | P<0.001 | 0.015 | 0.045 | P<0.001 | 0.273 | 0.001 | 0.371 | P<0.001 | - | | |
| m | 0.976 | 0.064 | 0.582 | 0.258 | P<0.001 | 0.082 | 0.398 | 0.495 | 0.292 | 0.011 | - | |
| n | 0.433 | 0.012 | 0.018 | 0.466 | P<0.001 | 0.094 | 0.856 | 0.737 | 0.009 | 0.003 | 0.474 | - |

Table 2: Post-hoc t-tests for MD (A) and $M\gamma$ (B). For each couple of regions, there are reported the obtained p-values. The cells corresponding to regions discriminated with high significancy ($P<0.001$) are coloured in gray. The statistical test was performed between the following selected ROIs: occipital lobe(a), temporal lobe(b), anterior pericallosal areas(c), genu(d) and splenium(e) of the corpus callosum, posterior pericallosal areas(f), frontal lobe(g), parietal lobe(h), thalamus(i), putamen(l), head of the caudate nucleus(m) and posterior limb of the internal capsule(n)